# Quantum Computing: Theoretical *versus* Practical Possibility

Gheorghe Sorin Paraoanu[a]

An intense effort is being made today to build a quantum computer. Instead of presenting what has been achieved, I invoke here analogies from the history of science in an attempt to glimpse what the future might hold. Quantum computing is possible in principle - there are no known laws of Nature that prevent it - yet scaling up the few qubits demonstrated so far has proven to be exceedingly difficult. While this could be regarded merely as a technological or practical impediment, I argue that this difficulty might be a symptom of new laws of physics waiting to be discovered. I also introduce a distinction between "strong" and "weak" emergentist positions. The former assumes that a critical value of a parameter exists (one that is most likely related to the complexity of the states involved) at which the quantum-mechanical description breaks down, in other words, that quantum mechanics will turn out to be an incomplete description of reality. The latter assumes that quantum mechanics will remain as a universally valid theory, but that the classical resources required to build a real quantum computer scale up with the number of qubits, which hints that a limiting principle is at work.

## Introduction

Quantum computing is a young discipline at the interface between computer science and quantum physics. It rests on the discovery that certain computational tasks, such as number factorization, period finding, and database search, could be speeded up if these problems are encoded in the states of systems called quantum bits (qubits) and in operations described by quantum mechanics. In a quantum processor, the qubit states can be superposed and entangled by sequences of externally controlled manipulations that result in unitary transformations (quantum gates). The effect of measurement on the qubits is described by the standard von Neumann projection postulate.[1]

It is somewhat counterintuitive that more efficient processing of information can be obtained by using quantum mechanics. Should not exactly the opposite be the case? Objects in quantum mechanics in general do not have well-defined properties, so any information associated with such properties should be diffuse and, owing to the randomness of the measurement process, unreliable to recover. After all, does not the Heisenberg uncertainty relation set a fundamental limit to our knowledge of physical variables? And are there not several no-go theorems (no-cloning,[2] no-reflection,[3] no-deleting[4]) indicating that some operations that are possible in a classical computer are forbidden in quantum mechanics? The clever trick of quantum computing is to bypass all of these objections by avoiding the encoding of classical bits of information directly into an observable or (as in classical Turing machines) into an "element of reality" of any type. A quantum processor is a just a system that is prepared in an initial state and evolves quantum mechanically under the application of a sequence of quantum gates; at the end of this evolution the state is measured. The result, which encodes the solution to a problem, is cast irreversibly--*alea jacta est*!--into the classical world. In other words, to perform a computation there does not have to be a one-to-one correspondence between the bit and the "it" at every step of the program: what matters are the correlations established in the final quantum state of the processor due to evolution[5]. As it turns out, quantum mechanics offers a way to produce stronger correlations --stronger than the correlations achievable by the above correspondence. This profound consequence follows because there cannot be any local realistic variables "hidden" behind the quantum formalism. The groundbreaking work of John Bell showed that no theory with such "hidden variables" will reproduce the predictions of quantum theory. The correlations between experimentally measured results (as predicted by quantum physics) are stronger than any theory of hidden variables satisfying conditions such as locality and realism,[6] which would be the classical-computing type of "it" to use for embedding the "bit." That is why, when examining the number of operations needed to solve a problem, a quantum computer could be more efficient than a classical one.

---

[a] Gheorghe Sorin Paraoanu received his Ph.D. degree in Physics at University of Illinois at Urbana-Champaign in 2001 and a M.Sc. degree in Philosophy at University of Bucharest in 1995. He currently is a senior scientist in the Low Temperature Laboratory at Aalto University, Finland, where his main scientific interests focus on degenerate quantum gases, mesoscopic superconducting devices, and the foundations of quantum mechanics.

I shall not attempt to give a historical account of the many achievements in the field of quantum computing. Instead, I shall try to see if something can be learned about what could be expected in the future based upon analogies from the history of science. Theoretical attempts to design such a quantum-computing machine have led to important insights into the fundamentals of quantum physics, but experimental progress has not been so spectacular, at least when counting the number of qubits operated. What can we expect in the future? The simplest answer is that these difficulties are inherent in a long journey and just require persistence and courage to overcome them. However, there might be another answer, namely, that the laws of Nature have never been tested for systems of such complexity, and as a result we will simply face the unexpected. Ironically, therefore, the failure to build a quantum-computing machine might translate into the discovery of new physics. I speculate that there are two ways in which this physics could manifest itself: either the quantum many-qubit state would collapse above a certain level of complexity, or there could be a limiting principle--a situation in which increasing the computing power would require more and more classical resources. I characterize the first attitude as "strong emergentism" and the second as "weak emergentism."

**Experimental Realizations**

Quantum computers do not yet exist. The experimental effort to design them is already more than ten years old, during which time the field has clearly advanced. Many physical systems have been studied as candidates for the magic qubit, including trapped ions, photons, superconducting circuits, and atomic nuclei in certain materials. The most important quantum algorithms have been demonstrated for few-qubit systems (less than ten). This looks like solid progress, and one might wonder what exactly stops physicists from declaring the problem solved, passing the task of adding more qubits to engineers, and moving on to other issues. The reason is that scaling up from a few-qubit demo processor to a machine that actually would be able to compete with classical computers in solving certain problems more efficiently requires not just a simple redesign of a chip. What exactly is the difficulty? It depends on the type of quantum processor. For example, with optical quantum processors the main issue is the smallness of the photon-photon interaction: Interaction is responsible for entanglement, which is needed in quantum computers, so to get more photons entangled (or higher probability of entangled pairs) one needs to ramp up the power of the lasers--a strategy that obviously runs rapidly into technical problems. In the NMR (nuclear magnetic resonance) version, the signal falls off exponentially with the number of qubits. In ion traps, various sources of decoherence and instabilities are the main constraining factors. With superconducting qubits, decoherence arising from electromagnetic degrees of freedom and impurities in the junctions is the biggest obstacle to scalability. When we add detectors into the picture, a new set of restrictions should be considered, arising from their limited efficiency.

Apart from the issue of decoherence - a significant phenomenon in the context of quantum computing - most theorists working in the field of quantum computing regard many of these problems as unfortunate technological limitations at present. They hope they can be surpassed, for example, through advances in material science--new technologies that would produce defect-free insulating barriers for Josephson junctions and more efficient detectors--or through new ideas in design (segmented ion traps, better microwave-engineered superconducting circuits). As for decoherence, if insufficiently alleviated by progress in microfabrication, materials, and design, then new ideas, such as the use of decoherence-free subspaces or topological quantum computing, could be used. Indeed, decoherence has been a constant worry since the beginning of the field of quantum computing,[7] but the appearance of error-correction codes and the quantum-threshold theorem has turned the table in favor of the optimists.[8]

In truth, in many cases, constant progress in science in a predictable direction does occur: Thomas S. Kuhn called this "normal science" as opposed to a revolutionary period in which a new paradigm emerges. In some cases, tedious, sustained work toward a goal pays off in new inventions, sometimes contrary to the expectations of outsiders, two examples being Thomas Edison's electric light bulb and the Wright brothers' airplane.

**Practical Hurdles as an Indicator of Fundamental Limits**

Is it obvious that something that looks perfectly plausible and in tune with state-of-the art science and technology can be done in principle? Not at all: history is full of cases in which common sense and accepted knowledge is defied by unexpected findings.

There is no compelling argument that building a quantum computer is impossible. The only disproof of quantum computing to date comes from the grinding of teeth of hundreds of graduate students and postdocs who are trying to make it work in practice.[9] As is often said, quantum computing is a theorist's dream and an experimentalist's nightmare. Moreover, a bit of contemplation on the nature of many-body quantum physics shows that nothing comes with a warranty when we talk about complex quantum systems such as those required for a quantum computer.[b]

I first observe that the Hilbert space used by a quantum processor is extremely large, which is still an understatement! For a system of only 1000 qubits (which would constitute a minimal requirement for a decent start-up quantum processor) the Hilbert space has dimensionality $2^{1000} = 10^{301}$. This number is so large that any human effort to grasp it intuitively is hopeless. One could argue that flipping a coin 1000 times produces the same number of potential combinations, which no one would find mysterious. But this argument is not valid. In the case of a coin, properties such as "heads" and "tails" are classical properties. Each time we flip the coin, one of these properties is actualized. In the case of a quantum computer, each time we perform a gate the properties do not actualize: they must remain as potentialities.[c]

One might then ask how we get around this problem in condensed-matter physics? After all, the number of quantum entities here, such as electrons in a metal, is of the order of $10^{23}$. The answer is that we use certain symmetries of the Hamiltonian resulting from regularities in the crystalline structure of the solid. For example, when calculating energy bands in a semiconductor, the electrons are assumed to move in a periodic potential because the atoms are placed in a regular lattice. The Hamiltonians involved in quantum computing are not of this type; they are time-dependent Hamiltonians acting for a finite time on individual qubits or on pairs of qubits. The resulting states are highly entangled, with a very different structure from those of the ground and excited states in condensed-matter systems. As their complexity increases, the time required to prepare and to characterize them increases as well. For example, to do quantum tomography on an ion-trap system of 8 qubits it took more than half a million measurements over more than 10 hours,[10] and analyzing the data is yet another big hurdle.[d] Not only is it impractical; it is blatantly impossible to do so for a number of qubits only one order of magnitude larger. Now enter any laboratory specializing in quantum computing: the "quantum processor" can be a device no larger than a few millimeters, but take a deep breath and look around. You see piles of generators, amplifiers, digitizers, lasers, and lenses, complicated vacuum machines and cryogenic instruments. The more qubits, the more classical devices needed to support them. Accounting for the classical resources needed to run a quantum computer is not straightforward,[11] and requires a serious dose of optimism to imagine all of this hardware multiplied by orders of magnitude and compressed into a laptop.

---

[b] I use the word "complexity" in a very general sense to cover, for example, high-complexity classes of problems (in the computational sense) and highly entangled many-particle states, including Schrödinger-cat states (superpositions of many-body states with distinct macroscopic properties). A mathematical formulation of quantum complexity would be valuable, but it is not clear what a generally useful one would be.

[c] This is also true for versions of quantum computing that use measurement to produce the gates, such as the one-way quantum computer. Here some properties are actualized, but some must remain potential (that is, some quantum superpositions have to be preserved).

In physics, we assume that the laws of Nature are infinitely precise within their domains of validity--lack of precision in experiments arises from our imperfect measuring apparatuses. In reality, however, this is only a working supposition: it is perfectly reasonable to imagine that the ultimate precision to which we can know the laws of Nature also depends on the total measurement resources that can be made available,[12] and these are limited by cosmological parameters such as the size of the Universe. Owing to resource limitations, it is impossible, for example, to violate the Bell inequalities for large spins above a certain limiting value.[13] In general, given that our resources for testing physical laws are always finite, and therefore that the precision in our measurements is inevitably limited, one wonders whether our belief in the exactness of the laws of Nature is an illusion.[14] Now, the reason we achieved the present miniaturization of classical computers is that the formidable overhead required to support and pamper matter and fields at the level of single quanta is not needed: classical computers process information by using incoherent, large electrical currents and potentials consisting of many electrons, which are stable and obtained in a relatively easy and reliable way. In a classical computer, dissipation works in our favor, resulting in the stabilization of the machine. In a quantum computer, we have to fight against dissipation for every single qubit in the processor.

The struggle to increase the number of qubits parallels perfectly the uphill battle to factorize numbers. Consider, for example, the latest record in factorization on classical computers,[15] that of a 200-digit number in 2005: an 80-CPU cluster was used and the calculation took three months. It is tempting to speculate that the source of the difficulty arises because the amount of (classical) information that can be potentially encoded in both a highly entangled state and in a large semiprime number is immense. In 1991, RSA Laboratories[e] established the RSA Factoring Challenge: they put forward a list of prime numbers to be factorized by the cryptography community.[16] The contest ended in 2007. Over a period of sixteen years, RSA primes with numbers of binary digits between roughly 300 and 700 have been factorized. Over the same span of time and over about the same period, systems of 3 to 7 qubits have been demonstrated in various physical systems. Based on the development of the field of quantum computing, we could predict that adding yet another qubit will be as difficult as factoring a number with roughly 800 digits. Classical factoring and quantum computing seem to go in parallel, and there is no reason to believe that the speed of quantum computing would suddenly increase.

---

[d] One can say that quantum tomography aims at extracting and mapping in a classical format all of the information contained in quantum states, while quantum computing attempts to extract only some information, namely, the solution to the problem. In this sense, quantum computing fares better. However, a quantum computer has to go through states that require a large amount of classical information to characterize. We do not have any warranty of how robust these states are and if they can be realized with reasonable resources.

**Two Scenarios**

Imagine that we struggle to build a quantum computer for another twenty years or so and we fail, in the sense that it has become clear that progress would be harder and harder as the number of qubits and quantum gates increases, in other words, that instead of some geometric progression, as in Moore's famous prediction for semiconductor-based computing technology, we will see not even a linear but a logarithmic increase in the number of qubits as a function of time. Call this the log-log-Moore law.

Suppose now that the experimentalists were confident that they fully understand all of the sources of relaxation, dephasing, instabilities, material defects, and the like, in their single-qubit and few-qubit systems, and that with the parameters they extracted from their experiments clear theoretical predictions are being made about what to expect when the number of qubits is doubled, tripled, and so on. We leave aside the case in which we discover that the states required for quantum computing are extremely unstable under external perturbations, or that, as a direct consequence of quantum mechanics, the inclusion of spurious higher-order effects such as co-tunneling or two-photon processes is in blatant contradiction with the intensities required to produce entanglement at a reasonably high rate and hence would shave off any advantage of quantum computing. In the latter case, we would be stuck in the same epistemological crevasse that Bohr opened over eight decades ago: there is no way to surpass quantum physics, which is the only scientifically valid way of doing science, that is, by declaring the reality of the properties of microphysical objects to be meaningless, and by tying all allowed statements about a microphysical system to a highly qualified specification of the measurement process.

This scenario probably would be very disappointing for most practitioners in the field of quantum computing, but physicists are good at dealing with frustration, and most likely the optimists and the very smart would continue to improve the design of the samples, to come up with new error correction ideas and fault-tolerant computing schemes, and to try other systems as qubits. But even then there could be good physics ahead. Assume that owing to some previously unknown decoherence effects, the states we want to produce in a quantum computer would collapse inevitably to other states, also described by the quantum-mechanical formalism. There are examples of such behavior: A GHZ (Greenberger-Horne-Zeilinger) state of say $N$ particles (which is a type of highly entangled state) collapses to a trivial state (the ground state of $N-1$ particles) under the loss (or ground-state projection) of a single particle. However, an entangled $W$ state of $N$ particles collapses to a $W$ state of $N-1$ particles under the same type of process. Further, two-mode Fock states (also called fragmented states) become phase-coherent only after detecting a few particles.[17] Thus, effects similar to chaotic behavior, or new phase transitions and symmetry-breaking states, could emerge, resulting in the system switching between complex many-body metastable states. It is also possible that these states will support some type of novel functionality or completely unexpected behavior. This is what happens, after all, in biology, which is a good illustration of the idea that the absence of large-scale regularities can lead to significant outcomes (us, for example).

---

[e] RSA Laboratories, located in Bedford, Massachusetts, is the Security Division of EMC (Egan Marino Corporation).

It also could be, however, that none of the above events occur, or that they occur to a certain degree and quantum-mechanical ways of avoiding them are proposed, but a quantum processor is still not in sight. I conjecture that there are two ways in which this could occur, corresponding to what I call "strong emergentism" and "weak emergentism."

The first scenario, "strong emergentism," corresponds to the situation in which a certain level of complexity is identified beyond which, for no known reason, all attempts to assemble a quantum processor and to run a quantum algorithm fail. We then would have discovered an experimental situation that cannot be described by quantum mechanics. Einstein would have been revenged, with revolutionary implications. There would be a fresh, experimentally-accessible physics beyond quantum mechanics to explore, and every physicist would want to unveil its mysteries.

This idea that there could be an unexplored physics beyond quantum mechanics has excited theorists for a long time, but where to search for it (for instance, below the Planck scale) has been beyond present technological possibilities. What is the likelihood that quantum mechanics would break down at such low energy scales? Has not quantum mechanics been thoroughly tested, and does it not work even for relatively large objects, as confirmed by quantum superpositions in SQUIDs and interfering macromolecules?[18] Such spectacular experiments have indeed been done: but large is not necessarily complex. In fact, for all of these experiments, a single collective observable (magnetic flux for SQUIDs and center-of-mass trajectory for molecule interference) has been identified that behaves quantum-mechanically. Surprisingly, however, the number of elementary particles that participate effectively in these superpositions is quite low: Although the magnetic fluxes and currents in flux qubits are truly macroscopic, the number of electrons that produce them is in the thousands and not, as one naively would expect, of the order of Avogadro's number.[19] Moreover, by other measures of these macroscopic superpositions, the "size" of the experimentally created Schrödinger cats in flux qubits can be even smaller, of the order of unity![20] In the end, the types of mesoscopic or macroscopic states for which quantum-mechanical behavior has been proven have been rather limited. The same is true for statistical properties: While it is often claimed that the Pauli exclusion principle has been tested with a precision of the order of $10^{-26}$,[21] it should be remembered that these tests have been performed on very specific states.

It would be preposterous for me to try to guess what the new physics would look like; it all depends on what experiments will reveal. One possibility could be that the quantum-to-classical transition will be clarified. For example, we could discover that a process of spontaneous collapse of complex many-body wavefunctions onto classical states exists. Lajos Diósi and Roger Penrose have argued that a process of spontaneous collapse of the wavefunction could be triggered by the gravitational self-energy.[22] In other words, merely the property of carrying mass (above a certain critical value) would suppress superpositions of an object in spatially distinct states. As noted above, in the case of a quantum computer the wavefunction could collapse above a critical level of complexity. Entire classes of states could coalesce in a way that standard quantum physics cannot explain, and possibly certain states can never be prepared.

The second scenario, "weak emergentism," corresponds to the situation in which it becomes harder and harder to scale the system up from the level of a few qubits, and no critical level of complexity is identified. A symptom of this would be the cluttering of your laboratory: more and more "classical" equipment would be needed to control the qubits and the interactions between them. Metaphorically speaking, this could be a reflection of the amount of information the state is carrying, which in turn is directly related to the preparation procedure, which is expressed in classical terms: it is a list of the operations to perform in the laboratory to obtain the desired initial state. In this sense, there is no guarantee that Nature has provided a physical basis to distinguish between any two possible lists of such operations, especially considering the vast number of their combinations.

Furthermore, to prepare a system means that a quantum level has become entangled with a classical system, which in turn is "measured" or at least belongs to the classical world. Now, if the quantum state is complex, the risk is that there are too many classical systems to which one quantum state is entangled.[f] Therefore, each many-body state has to be associated with a classical set of objects or preparation procedures,[g] and the classical states of the surrounding

---

[f] This argument is weakened, of course, because to prepare a quantum state it is not really necessary, as I noted earlier, to have a one-to-one correspondence between quantum and classical states. Quantum states also can be prepared by having quantum systems interact with each other. However, to do this in a controllable way, precise classical-level manipulations of fields, potentials, and the like, are still required.

[g] One can argue that states such as BCS (Bardeen-Cooper-Schrieffer) states look rather complex, yet there is no problem to obtain them because they are the ground states of simple Hamiltonians. But BCS states have a simple underlying symmetry: they are constructed from adding pairs of electrons with opposite

objects are mapped onto the state of the quantum computer. That also happens in classical computers, but the difference is that we claim that the quantum computer achieves a more efficient organization of this information. While this might be the case, but it also could be that a rather large amount of "spurious" information must be necessarily left behind in the environment (or is required to begin with). And the classical resources needed to account for this information can be quite large, a point that is usually ignored in quantum-computing science: Only the number of gates is counted, which is compared to the equivalent number of gates required in a classical computer. Not much investigated is what is behind a quantum gate; it is simply assumed that the technology will be sufficiently advanced to easily realize these operations.

In this second scenario, while we would fail to produce something potentially useful for society, we would have put our finger on something deep that requires explanation. Why would Nature so fiercely oppose factorizing numbers? It is fair to assume that *Der Alte* has little interest in hacking into our bank accounts or in finding the latest military strategic game. What then is there in Nature that would prevent us from doing calculations for fast-searching, for example? Is our existence in this Universe compatible with the possibility of performing certain mathematical tasks? Does the anthropic principle rule out quantum computing?

Number theory is one of the most difficult branches of mathematics, yet a result like Fermat's last theorem does not require more than elementary-school mathematics to formulate. We do not know the answer to simple questions, such as the distribution of primes among the integers - Riemann's hypothesis has not yet been proven. An outsider might think that number theory should be the primary mathematical tool for a physicist: after all, numbers are what we get from experiments. But concepts such as prime numbers, which form the backbone of number theory, have little relevance for the natural sciences: The numbers that come out of physics experiments are a string of rational numbers, and mathematical questions like countability or the density of rational numbers in the set of reals are either irrelevant or completely buried under instrumental errors. Physics describes the continuous dynamics of point objects and fields in space and time, and therefore the natural tools of the physicist should be calculus, differential geometry, group theory, and linear algebra. However, surprising connections between number theory and physics have been discovered.[23] Perhaps we are just scratching the top of the iceberg, because these connections seem to occur precisely in areas such as renormalizable field theories and low-dimensional field theories, where (despite their great experimental success) we are least confident that their concepts are optimal. Indeed, future physicists probably would regard summing over Feynman diagrams as conceptually awkward and cumbersome as today's physicists regard combining motions of epicycles to get planetary orbits in our solar system.

---

momentum and spin to vacuum. Even so, BCS states actually are not so easily available: setting aside the low-temperature resources required, such states appear only in a few metals.

It is truly astonishing how little we know about such deceptively simple concepts as numbers. We do not really understand the origin of the building blocks of arithmetic, the prime numbers. We have discovered that arithmetic cannot be axiomatized completely and consistently, and we have found deep connections between most of the other branches of mathematics and physics. We also seem to have uncovered some ways in which quantum mechanics is connected to one of the most elementary mathematical concepts imaginable, that of a number. Counting is surely the first mathematical tool we learn as children, and we are amazed to discover that much of the construction of the Universe relies on this skill. Perhaps, after all, in the beginning was the Number.

## A Limiting Principle

The notion that there are limiting principles in science is not without precedent. At the end of the nineteenth century, Ludwig Boltzmann was firmly convinced of the existence of atoms, but the technology necessary to isolate an individual atom lay in the future. In mathematics, David Hilbert's program of axiomatization must have looked challenging but feasible to Bertrand Russell and Alfred North Whitehead, but then Kurt Gödel's incompleteness theorem undercut it. Again, there were good reasons to suspect that it would be impossible to cool an object to lower and lower temperatures. There are many examples of phenomena that produce heat naturally to hundreds and thousands of degrees Celsius, but not many that produce below-zero Celsius temperatures. Cold is difficult to create. In the biological world, animals have developed sophisticated means to produce heat at the cellular level, but the trick they use to cool off is rudimentary: sweating. The instruments we use to measure temperature offer another hint: It is difficult to create a simple thermometer that works at very low temperatures, since most fluids freeze. We know, of course, that the fundamental limit is zero degrees Kelvin, which is unattainable.[h]

In the case of quantum complexity, everything seems to indicate that its limit is close to our present technological capabilities. We might reach a universal fundamental limit that would combine thermodynamical and information-theoretical concepts. The truly desirable forms of quantum computing might be possible only asymptotically, by either requiring such extensive resources that even relativistic effects become important, or by being confined to such a small space that virtual excitations begin to play a role. Hilbert space indicates what is possible according to what we know about a number of particular cases, but we have no warranty that all of these vast arrays of possibilities are physical. I imagine this limiting principle in analogy to the second law of thermodynamics. Suppose that we had studied isothermal processes at a time before the formulation of this law in the middle of the nineteenth century. A consequence of Boyle's law is that an ideal gas in contact with a heat reservoir expands isothermally and produces work, so we might have confidently predicted that a machine could be built that would transform all of the energy in a heat reservoir into work. The only worry would be that in the isothermal expansion of a gas its volume increases. Somehow, the functioning of the machine has left a mark on the environment (similar to the current difficulties with decoherence and the increase in classical resources). To have a truly functional machine requires other processes to bring the volume of the gas back to its initial value. We are thus led to discover that a cold reservoir is also required to built a machine that converts heat into work. In an even closer analogy to my claim that certain states might be physically unattainable in a quantum computer, the Carathéodory formulation of the second law shows that thermodynamic states that cannot be reached by adiabatic processes always exist in the vicinity of an initial state. Thus, a limiting principle exists in Nature that prevents us from building a heat engine that is 100% efficient - complete conversion of heat into mechanical work is forbidden (most machines, of course, are far less efficient).

---

[h] The record for the lowest temperature ever produced in a metal belongs to the Low Temperature Laboratory at Aalto University in Finland: 100 picokelvin ($10^{-12}$ Kelvin). At the Massachusetts Institute of Technology, temperatures of the order of 500 picokelvin have been obtained recently in a different system, a Bose-Einstein condensate.

A similar limiting principle could hold for information processing, one that prevents us to do certain fast calculations and outperform classical computers. Recall that quantum algorithms seem to offer a more efficient way to process correlations between logical entities without requiring them to be represented by (separate) physical entities. Thus, while running a quantum algorithm, the amount of classical information generated by the recording of intermediate results and dissipation is much smaller than that required by a classical computer performing the same task. Ideally, this difference should be zero, meaning that the states in a quantum processor do not decay, get entangled with the states of other nearby devices (those used for measurement, trapping, and so on), or do anything whatsoever that would broadcast information to the rest of the Universe. This extraneous classical information, plus the information required to create and manipulate the state (encoded even in the laboratory instruments), could play the role of the heat that must be dissipated to extract work from a heat reservoir.

Further, factorization can be viewed as the mathematical equivalent of splitting matter: prime numbers are the "atoms" that comprise any integer number. Not being able to factorize numbers efficiently could be a symptom of the more general idea that it is not always possible to break down complex, large-scale entities into distinct, "elementary" units. That is, it hints at a failure of reductionism already at the mathematical level. "More is different," as P.W. Anderson put it in his famous 1972 essay.[24] Indeed, although things are made of parts, given the large numbers involved it would be impossible in practice to express the higher-order, emergent laws in terms of elementary components. If we accept this, then the reason that gravity is so notoriously difficult to reconcile with quantum mechanics becomes more understandable: The general theory of relativity is an "intensely classical" theory, requiring concepts such as clocks, measuring rods, and so on, which (unlike quantum-mechanical entities) have intrinsic properties. This conceptual tension already exists at the level of marrying quantum physics and special relativity, even if physicists generally believe that is has been resolved successfully in relativistic field theory.[25]

But if reducing higher-level laws to lower-level laws would require exponentially increasing resources (in the number of components), then it could be that a true epistemological gap would exist between "elementary" and "emergent" laws:  We not only will never find a theory of everything but the very concept of such a theory will be proven to be  inconsistent. In other words, our ability to completely classify reality into elementary entities is impaired when we consider complex objects, which have no underlying symmetry. Quantum computing can then be regarded as a particular attempt to bypass this gap, by assuming full control over the emergent properties of a complex system by manipulating its individual components.

## Conclusions

A distinction always can be made between the theoretical and practical possibility of building a device. The failure to build a device in practice is usually taken to indicate a lack of technical skill or knowledge on the part of the builder. I have argued that in the case of quantum computing we should amend this statement and seriously consider alternatives. I have identified two logical positions, "strong" and "weak" emergentism. In the former, we would find that quantum physics does not offer a complete description of the physics of complex quantum states such as those involved in the quantum processing of information. In the latter, we would be forced to postulate a limiting information-thermodynamical principle that forbids us to perform highly complex calculations efficiently.

## Acknowledgments

My research for this paper began under a John Templeton Fellowship, which allowed me to spend the summer of 2009 at the Institute of Quantum Optics and Quantum Information of the University of Vienna.  I especially thank my hosts, Professor Anton Zeilinger and Professor Markus Aspelmeyer, who made this visit possible, and for many enlightening discussions with them and other scientists in the Institute.  I am solely responsible, of course, for the (probably controversial) views I express.   I also thank the Academy of Finland for financial support through Academy Research Fellowship 00857 and Projects 129896, 118122, 135135, and 141559.  Finally, I thank an anonymous referee for helpful comments, and Roger H. Stuewer for his editorial work on my paper.

Low Temperature Laboratory
Aalto University
P. O. Box 15100
FI-00076 AALTO, Finland
e-mail: sorin.paraoanu@aalto.fi